\documentclass[a4paper,12pt]{article}
\usepackage{graphicx}
\usepackage{epsfig}
\usepackage{amsmath}

\newcommand{\be}{\begin{equation}}
\newcommand{\ee}{\end{equation}}
\begin{document}

\title{Strong shock in a uniformly expanding medium }

\author{G.S. Bisnovatyi-Kogan
\thanks{ Space Research Institute,
Russian Academy of Sciences, Moscow, Russia;
National Research Nuclear University MEPhI,
 Moscow, Russia;  and ICRANet, CBPF, Rio de Janeiro, Brazil.}}
\date{}
\maketitle

\begin{abstract}
\noindent Propagation of the strong shock in the flat expanding Friedman universe is investigated using methods of
dimension and similarity. Exact analytic solution of self-similar equations is obtained, determining
dependences of the radius and velocity of the shock wave on time and radius.
It is obtained, that in the expanding medium the velocity of shock
decreases as $\sim t^{-1/5}$, what is  slower than the shock velocity in the static uniform medium $\sim t^{-3/5}$.
The radius of the shock wave in the expanding self-gravitating medium
increases $\sim t^{4/5}$, more rapidly than the shock wave radius in the uniform non-gravitating medium $\sim t^{2/5}$.
So, the shock propagates in the direction of decreasing density with larger speed, that in the static medium,
 due to accelerating action of the decreasing density, even in the presence of a self-gravitation.

\end{abstract}


\section{Introduction}

At early stages of star and galaxy formation we may expect strong explosions at last stages of evolution
of very massive primordial stars, which enrich the matter with heavy elements. Detection of heavy elements at
red shifts up to $z\sim 10$ from GRB observations (GRB090423 at $z\approx  8.2$,  GRB120923A at  $z\approx 8.5$,
GRB090429B with a photo-$z \approx 9.4$) \cite{lzgrb} make plausible this suggestion.
Propagation of a strong shock in the static uniform media was investigated by many authors \cite{stanyuk},
\cite{taylor}, but the finite analytic self-similar solution was obtained in \cite{sedov}.
Here we obtain a self-similar solution for the strong shock propagating through the uniform expanding
 media, corresponding to the Friedman solution for the flat universe \cite{znuniv}.

\section{Self-similar solution for a strong shock propagation in a uniform static medium }

Let us first describe the self-similar solution of Sedov \cite{sedov}, following \cite{llhydro}.
Hydrodynamic equations in spherically symmetric isentropic case are written as

\be
 \frac{\partial v}{\partial t}+v \frac{\partial v}{\partial r}=-\frac{1}{\rho} \frac{\partial p}{\partial r},
 \quad \frac{\partial \rho}{\partial t}+ \frac{\partial \rho v}{\partial r}+\frac{2\rho v}{r}=0,
 \quad
\left(\frac{\partial }{\partial t}+v \frac{\partial }{\partial r}\right)\ln{\frac{p}{\rho^\gamma}}=0.
\label{eq1}
\ee
Here $v$ is a gas velocity in the laboratory coordinate system, $\rho$ is the gas density, $p$ is the gas pressure.
Defining the variables with  subscripts  "1" and "2" corresponding to the values in front and behind the shock wave,
we may write the relations on the shock,
moving through the static media $(v_1=0)$ in the form  \cite{sedov1}

$$
 v_2=\frac{2}{\gamma+1}u_1\left(1-\frac{c_1^2}{u_1^2}\right),
 $$
 \be
\rho_2=\frac{\gamma+1}{\gamma-1}\rho_1\left(1+\frac{2}{\gamma-1}\frac{c_1^2}{u_1^2}\right)^{-1},
\label{eq2}
\ee
$$
p_2=\frac{2}{\gamma+1}\rho_1 u_1^2\left(1-\frac{\gamma-1}{2\gamma}\frac{c_1^2}{u_1^2}\right).
$$
Here $u_1$ is the shock velocity relative to the unmoving gas in front ($v_1=0$), $\gamma$ is the
adiabatic power, $c_1$ is the sound velocity in the static gas, $c_1^2=\gamma\left(\frac{p_1}{\rho_1}\right)$.
In the strong shock the shock velocity is much larger than the sound velocity in the undisturbed gas,
$u_1\gg c_1$, and $p_2\gg \frac{\gamma+1}{\gamma-1}p_1$, so we have from (\ref{eq2})

\be
 v_2=\frac{2}{\gamma+1}u_1,\,\, \rho_2=\frac{\gamma+1}{\gamma-1}\rho_1,\,\,p_2=\frac{2}{\gamma+1}\rho_1 u_1^2,
 \,\,\, c_2^2=\frac{2\gamma(\gamma-1)}{(\gamma+1)^2}u_1^2.
 \label{eq3}
\ee
There are only two parameters in the case of the strong shock: the density of the unperturbed gas $\rho_1$.
and the total energy of the explosion $E$. From these parameters, together with the independent variables
$t,\, r$, it is possible to construct only one non-dimensional combination $r(\rho_1/Et^2)^{1/5}$, what
means that the problem has a self-similar solution. A position of the shock in the self-similar solution
should correspond to fixed value of the self-similar coordinate, so that the distance of the shock
 to the center $R$ may be written as

 \be
 R=\beta\left(\frac{E t^2}{\rho_1}\right)^{1/5},
 \label{eq4}
 \ee
 where $\beta$ is a number depending only on the adiabatic power $\gamma$. The velocity of the shock $u_1$ in the
 laboratory frame, relative to the unperturbed gas is defined as

 \be
 u_1=\frac{dR}{dt}=\frac{2R}{5t}=\frac{2\beta E^{1/5}}{5\rho_1^{1/5} t^{3/5}}.
\label{eq5}
 \ee
So, the velocity of the shock $u_1$, the velocity of the matter behind the shock $v_2$,
moving through the constant density medium  are decreasing with time
  $\sim t^{-3/5}$, and pressure behind the shock $p_2 \sim t^{-6/5}$.

 Let us use a sound velocity variable $c^2=\gamma p/\rho$ instead of a pressure $p$.
 To find a full solution of the problem introduce non-dimensional variables
behind the shock $(V,G,Z)$ as

 \be
 v=\frac{2r}{5t} V,\,\,\, \rho=\rho_1 G,\,\,\, c^2=\frac{4r^2}{25 t^2} Z.
 \label{eq6}
 \ee
 depending on the self-similar variable $\xi$, defined as

 \be
\xi= \frac{r}{R(t)}= \frac{r}{\beta}\left(\frac{\rho_1}{Et^2}\right)^{1/5}.
 \label{eq7}
 \ee
The conditions (\ref{eq3}) on the strong shock at $r=R$, $\xi=1$, in non-dimensional variables (\ref{eq6})
are written as

\be
V(1)=\frac{2}{\gamma+1}, \,\,\, G(1)=\frac{\gamma+1}{\gamma-1},\,\,\, Z(1)=\frac{2\gamma (\gamma-1)}{(\gamma+1)^2}.
\label{eq8}
\ee
In non-dimensional variables (\ref{eq6}) the original system (\ref{eq1}) is written as

\be
Z\left(\frac{d\ln Z}{d\ln\xi}+\frac{d\ln G}{d\ln\xi}+2\right)+\gamma(V-1)\frac{d V}{d\ln\xi}=\gamma V(\frac{5}{2}-V),
\label{eq9}
\ee

\be
\frac{d V}{d\ln\xi}-(1-V)\frac{d\ln G}{d\ln\xi}=-3V,
\label{eq10}
\ee
\be
\frac{d\ln Z}{d\ln\xi}-(\gamma-1)\frac{d\ln G}{d\ln\xi}=-\frac{5-2V}{1-V}.
\label{eq11}
\ee
Here we used relations

\be
\frac{\partial\xi}{\partial t}|_r=-\frac{2\xi}{5t},\quad  \frac{\partial\xi}{\partial r}|_t=\frac{\xi}{r}.
\label{eq12}
\ee

\subsection{Energy conservation integral}

The full analytic solution of the problem was obtained by L.I.Sedov, who found the expression for the full energy
integral $E$, what permitted to omit the equation (\ref{eq9}), see \cite{sedov}. The conservation of full energy
in this problem is a result of the propagation through
the medium with zero pressure, what means zero
external work. Due to self-similarity there should be conservation of the total energy inside any sphere
of smaller radius, which
increases with time according to the low $\xi=const$, with any value of the $const$. Points on the sphere of the radius $r$ expanding
according this low, have a velocity $v_n=\frac{2r}{5t}=u_1\frac{r}{R}$.
A simple physical derivation of the energy conservation low was given in the book \cite{llhydro}.

The change of the energy of the spherical region with a radius $r$, during the time $dt$,
is determined by its loss due to the flux of the energy $dq_-$ carried away by the matter, expanding with the velocity $v$.
Using Bernoulli integral, the flux through the spherical surface with radius $r$ during the time $dt$ is written as

\be
dq_- = dt\,4\pi r^2\rho v (w+\frac{v^2}{2})
\label{eq13}
\ee
The  gain of the energy $dq_+$ is due to increase of the volume of this sphere during the time $dt$
by the value ($dt\,v_n 4\pi r^2$), containing the energy

\be
dq_+ = dt\,4\pi r^2\rho v_n (\varepsilon+\frac{v^2}{2}),\quad w=\varepsilon+\frac{p}{\rho}.
\label{eq14}
\ee
From the equality $q_-=q_+$ we have

\be
v(w+\frac{v^2}{2})=v_n (\varepsilon+\frac{v^2}{2}).
\label{eq15}
\ee
From thermodynamic relations we have \cite{llstat}

\be
\varepsilon=\frac{c^2}{\gamma(\gamma-1)},\quad w=\frac{c^2}{\gamma-1},
\label{eq16}
\ee
so that (\ref{eq15}) is written as

\be
v\left(\frac{c^2}{\gamma-1}+\frac{v^2}{2}\right)=v_n (\frac{c^2}{\gamma(\gamma-1)}+\frac{v^2}{2}).
\label{eq17}
\ee
It follows from (\ref{eq5}),(\ref{eq6}) the relation $\frac{v}{v_n}=V$, and we have from (\ref{eq17})

\be
\frac{c^2}{\gamma-1}(V-\frac{1}{\gamma})=\frac{v^2}{2}(1-V)
\label{eq18}
\ee
Using (\ref{eq6}) we write the energy conservation low in the form \cite{sedov1},\cite{llhydro}

\be
Z=\frac{\gamma(\gamma-1)(1-V)}{2(\gamma V-1)}V^2.
\label{eq19}
\ee
Evidently, that at the shock $r=R$, $\xi=1$, with $V(1)$ and $Z(1)$ from (\ref{eq8}), the
energy integral (\ref{eq19}) becomes an identity.
At existence of the energy integral (\ref{eq19}) we may use only two differential equations
(\ref{eq10}) and (\ref{eq11}) for finding a solution of the problem. It was shown in \cite{sedov},
that this solution may be obtained in the analytical form.

\subsection{Exact solution}

Following \cite{sedov},\cite{sedov1}, we look for an analytic solution of (\ref{eq10}),(\ref{eq11}), where
the variable $Z$ is determined by the algebraic relation (\ref{eq19}).
Excluding $\frac{d\ln G}{d\ln\xi}$ from (\ref{eq10}),(\ref{eq11}), we obtain the equation

\be
(\gamma-1)\frac{dV}{d\ln\xi}-(1-V)\frac{d\ln Z}{d\ln\xi}=5-3\gamma V+V.
\label{eq20}
\ee
From (\ref{eq19}) we find
\be
\frac{d\ln Z}{d\ln\xi}=-\frac{1}{1-V}\,\frac{dV}{d\ln\xi}+2\frac{1}{V}\,\frac{dV}{d\ln\xi}-\frac{\gamma}{\gamma V-1}\,\frac{dV}{d\ln\xi}.
\label{eq21}
\ee
After substituting in (\ref{eq20}) we find the following equation for the variable $V$

\be
\left(\gamma+1-\frac{2}{V}+\frac{\gamma-1}{\gamma V-1}\right)\frac{d V}{d\ln \xi}=5-3\gamma V +V.
\label{eq22}
\ee
Using relations
\be
-\,\frac{2}{V(5-3\gamma V+V)}=-\frac{2}{5V}+\frac{2}{5}\,\frac{1-3\gamma}{5-3\gamma V+V}, \quad
\label{eq23}
\ee
$$\frac{\gamma-1}{(\gamma V-1)(5-3\gamma V+V)}=\frac{\gamma-1}{2\gamma+1}\,\frac{\gamma}{\gamma V-1}
+\frac{\gamma-1}{2\gamma+1}\,\frac{3\gamma-1}{5-3\gamma V+V},
$$
we obtain a solution of (\ref{eq22}) in the form

\be
\frac{\gamma+1}{1-3\gamma}\,\ln (5-3\gamma V+V)-\frac{2}{5}\,\ln V+\frac{2}{5}\,\ln(5-3\gamma V+V)
\label{eq24}
\ee
$$ +\frac{\gamma-1}{2\gamma+1}\,\ln(\gamma V-1)-\frac{\gamma-1}{2\gamma+1}\,\ln (5-3\gamma V+V)
=\ln\xi+{\rm Const}.
$$
Using the boundary condition for $\xi=1$ from (\ref{eq8}), we obtain the solution for $V(\xi)$ as \cite{llhydro}

\be
\left(\frac{\gamma+1}{2}V\right)^{-2}\,\left[\frac{\gamma+1}{\gamma-1}(\gamma V-1)\right]^{\nu_1}\,
\left[\frac{\gamma+1}{7-\gamma}(5-3\gamma V+V)\right]^{\nu_2}=\xi^5,
\label{eq25}
\ee
with

\be
\nu_1=5\frac{\gamma-1}{2\gamma+1}, \quad \nu_2=-\frac{13\gamma^2-7\gamma+12}{(3\gamma-1)(2\gamma+1)}.
\label{eq26}
\ee
For finding a solution for $G(\xi)$ we write the equations (\ref{eq10}) and (\ref{eq22}) in the form

\be
1-(1-V)\frac{d \ln G}{d V}=-3V \frac{d \ln\xi}{d V},
\label{eq27}
\ee
\be
\frac{d \ln\xi}{d V}=\frac{\gamma+1-\frac{2}{V}+\frac{\gamma-1}{\gamma V-1}}{5-3\gamma V +V}.
\label{eq28}
\ee
The equation for $G(V)$ is obtained from these equations in the form

\be
1-(1-V)\frac{d \ln G}{d V}=-3V\frac{\gamma+1-\frac{2}{V}+\frac{\gamma-1}{\gamma V-1}}{5-3\gamma V +V},
\label{eq29}
\ee
which is reduces to

\be
(1-V)\frac{d \ln G}{d V}=\frac{2(1+2V)+3\frac{1-V}{\gamma V-1}}{5-3\gamma V +V}.
\label{eq30}
\ee
Using relations

\be
\frac{2+4V}{(5-3\gamma V +V)(1-V)}=\frac{2}{2-\gamma}\,
\frac{1}{1-V}-2\frac{\gamma+3}{2-\gamma}\,\frac{1}{5-3\gamma V +V},
\label{eq31}
\ee
$$
\frac{3}{(\gamma V-1)(5-3\gamma V +V)}=\frac{3\gamma}{2\gamma+1}\,\frac{1}{\gamma V-1}+3\frac{3\gamma-1}{2\gamma+1}\,
\frac{1}{5-3\gamma V +V},
$$
we obtain the solution of (\ref{eq30}) in the form

\be
\ln G=-\frac{2}{2-\gamma}\ln(1-V)+\frac{3}{2\gamma+1}\ln(\gamma V-1)
\label{eq32}
\ee
$$+\frac{13\gamma^2-7\gamma+12}{(2-\gamma)(3\gamma-1)(2\gamma+1)}\ln(5-3\gamma V +V)+{\rm\bf const}.
$$
The {\bf const} is found from the boundary conditions (\ref{eq8}), and finally we obtain the solution for $G(V)$ in the form

\be
\label{eq32b}
G(V)=\frac{\gamma+1}{\gamma-1}\left[\frac{\gamma+1}{\gamma-1}(1-V)\right]^{\sigma_1}
\left[\frac{\gamma+1}{\gamma-1}(\gamma V-1)\right]^{\sigma_2}
\ee
$$
\times \left[\frac{\gamma+1}{7-\gamma}(5-3\gamma V+V)\right]^{\sigma_3},
$$
where

\be
\label{eq32c}
\sigma_1=-\frac{2}{2-\gamma},\,\,\,
\sigma_2=\frac{3}{2\gamma+1},\,\,\,
\sigma_3=\frac{13\gamma^2-7\gamma+12}{(2-\gamma)(3\gamma-1)(2\gamma+1)}.
\ee
The function $Z(V)$ is determined by the integral (\ref{eq19}).

The constant $\beta$ in the definition of the non-dimensional radius $\xi$ in (\ref{eq7}) is obtained from the
energy integral $E$ for the known energy of the explosion. The conserving value of the energy behind the shock is determined as
\cite{llhydro}

\be
\label{eq32d}
E=\int_0^{R(t)} \rho\left[\frac{v^2}{2}+\frac{c^2}{\gamma(\gamma-1)}\right]4\pi r^2 dr.
\ee
In non-dimensional variables (\ref{eq6}) this relation reduces to the equation for the constant $\beta$

\be
\label{eq32e}
1=\beta^5\frac{16\pi}{25}\int_0^1 G\left[\frac{V^2}{2}+\frac{Z}{\gamma(\gamma-1)}\right]\xi^4 d\xi
\ee
The value of $\beta$, following from this relation is of the order of unity \cite{llhydro},
for $\gamma=\frac{7}{5}$ we have $\beta=1.033$.

\section{Strong shock in a uniform expanding medium }

Let us consider a uniformly expanding medium,
corresponding to the flat expanding Friedman universe \cite{znuniv}, with
a density $\rho_1$, and expansion velocity $v_1$ as

\be
\rho_1=\delta/t^2,\qquad v_1=2r/3t.
\label{eq20a}
\ee
This is an exact solution for a uniform expanding self-gravitating medium with velocity tending to
zero at time infinity. The equations in spherical coordinates in this case are written as

\be
 \frac{\partial v}{\partial t}+v \frac{\partial v}{\partial r}=-\frac{1}{\rho} \frac{\partial p}{\partial r}-\frac{G_g m}{r^2},
 \quad \frac{\partial \rho}{\partial t}+ \frac{\partial \rho v}{\partial r}+\frac{2\rho v}{r}=0,
\label{eq1a}
\ee
$$
\quad
\left(\frac{\partial }{\partial t}+v \frac{\partial }{\partial r}\right)\ln{\frac{p}{\rho^\gamma}}=0,
\quad \frac{\partial m}{\partial r}={4\pi}\rho r^2.
$$
The solution (\ref{eq20a}) satisfies equations (\ref{eq1a}) at ($G_g$ - gravitational constant)

\be
\delta=\frac{1}{6\pi G_g},\quad \rho_1=\frac{1}{6\pi G_g t^2},\quad \frac{G_g m}{r^2}=\frac{2}{9}\frac{r}{t^2}.
\label{eq1b}
\ee
The newtonian solution for the flat expanding universe is valid physically in the region where
$v_1\ll c_{\rm light}$, $c\ll c_{\rm light}$.
For the case of a point explosion with the energy $E$,
the number of parameters  is the same as in the previous static medium
($\delta, \,\,\, E$), therefore
we look for a self-similar solution for the case of a strong shock motion.
 The non-dimensional combination in the case of a uniformly
expanding medium is written as $r(\delta/Et^4)^{1/5}$. A position of the shock in the self-similar solution
should correspond to the fixed value of the self-similar coordinate, so that the distance of the shock
 to the center $R$ may be written as

 \be
 R=\beta\left(\frac{E t^4}{\delta}\right)^{1/5},
 \label{eq21a}
 \ee
 where $\beta$ is a number depending only on the adiabatic power $\gamma$. The velocity of the shock $u$ in the
 laboratory frame, for the unperturbed expanding gas is defined as

\be
 u=\frac{dR}{dt}=\frac{4R}{5t}=\frac{4\beta E^{1/5}}{5\delta^{1/5} t^{1/5}}.
\label{eq22a}
 \ee
The velocity of the shock $u$, the velocity of the matter behind the shock $v_2$, moving through the uniformly expanding
medium (\ref{eq20a}), are decreasing with time
$\sim t^{-1/5}$, and the pressure behind the shock $p_2$ is decreasing $\sim t^{-2/5}$,
what is considerably slower than in the case of the constant density
medium. It happens due to the fact, that the background density is decreasing with time, and the resistance to the shock
propagation is decreasing also.

The Euler equations of motion (\ref{eq1}) are valid in this case, but the conditions on the strong shock discontinuity
should be written, with account of the expansion in the laboratory frame as

\be
 v_2=\frac{2}{\gamma+1}u+\frac{\gamma-1}{\gamma+1}v^{sh}_1,\,\, \rho_2=\frac{\gamma+1}{\gamma-1}\rho_1,\,\,
 \label{eq23a}
 \ee
$$ p_2=\frac{2}{\gamma+1}\rho_1 (u-v^{sh}_1)^2,\,\,
 c_2^2=\frac{2\gamma(\gamma-1)}{(\gamma+1)^2}(u-v^{sh}_1)^2.
 $$
Here $v_1^{sh}=\frac{2R}{3t}$ is the unperturbed expansion velocity on the shock level. We introduce non-dimensional variables behind the shock as

 \be
 v=\frac{4r}{5t} V,\,\,\, \rho=\frac{\delta}{t^2} G,\,\,\, c^2=\frac{16r^2}{25 t^2} Z,\,\,\,m=\frac{4\pi}{3}\rho r^3 M=\frac{4\pi}{3}\frac{r^3}{t^2}\delta G\,M,
 \label{eq24a}
 \ee
depending on the self-similar variable $\xi$, defined as

 \be
\xi= \frac{r}{R(t)}= \frac{r}{\beta}\left(\frac{\delta}{Et^4}\right)^{1/5}.
 \label{eq25a}
 \ee
The conditions (\ref{eq23a}) on the strong shock at $r=R$, $\xi=1$, in non-dimensional variables (\ref{eq24a})
are written as

\be
V(1)=\frac{5\gamma+7}{6(\gamma+1)}
, \,\,\, G(1)=\frac{\gamma+1}{\gamma-1},\,\,\,
Z(1)=\frac{\gamma (\gamma-1)}{18(\gamma+1)^2},\,\,\,M(1)=1.
\label{eq26a}
\ee
In non-dimensional variables (\ref{eq24a}) the original system (\ref{eq1}) is written as

\be
Z\left(\frac{d\ln Z}{d\ln\xi}+\frac{d\ln G}{d\ln\xi}+2\right)+\gamma(V-1)\frac{d V}{d\ln\xi}
=\gamma V(\frac{5}{4}-V)-\frac{25}{72}\gamma G M,
\label{eq27a}
\ee
\be
\frac{d V}{d\ln\xi}-(1-V)\frac{d\ln G}{d\ln\xi}=-3V+\frac{5}{2},
\label{eq28a}
\ee
\be
\frac{d\ln Z}{d\ln\xi}-(\gamma-1)\frac{d\ln G}{d\ln\xi}=-\frac{5-2V-\frac{5}{2}\gamma}{1-V}.
\label{eq29a}
\ee

\be
\frac{d\ln M}{d\ln\xi}+\frac{d\ln G}{d\ln\xi}=3\left(\frac{1}{M}-1\right).
\label{eq29b}
\ee

Here we used relations

\be
\frac{\partial\xi}{\partial t}|_r=-\frac{4\xi}{5t},\quad  \frac{\partial\xi}{\partial r}|_t=\frac{\xi}{r}.
\label{eq30a}
\ee

\subsection{Construction of the first integral of the problem}

 The system of three equations (\ref{eq27a})-(\ref{eq29a}), describing the self-similar solution of the problem,
 has a first integral, similar to the energy conservation integral in the case of the static background medium
 (\ref{eq19}). To find this integral, we consider a situation in the coordinate system, where the background
 medium is locally static on the shock at $r=R(t)$. In this non-inertial coordinate system it is possible to
 construct the combination of functions, representing the first integral. We cannot call this integral as an
 energy integral, because the the system at which it has the same form as (\ref{eq19}) is not inertial.
  Due to self-similarity there should be conservation of this integral inside any sphere
of smaller radius $r\leq R(t)$. The  frame  which is locally comoving at $r=R$, has a velocity $v_1^{sh}=\frac{2R}{3t}$
relative to the laboratory frame, and at $r\leq R(t)$ we should use instead, due to self-similarity, the velocity
$v_1^{sh}\frac{r}{R}=\frac{2r}{3t}$, like in the case of the static background.
This velocity should be subtracted from all velocities in the expression for the first integral, which
has the form (\ref{eq17}). At $r\leq R(t)$ the velocity  $v_n=u\frac{r}{R}=\frac{4r}{5t}$.   We have than

\be
(v-v_1^{sh}\frac{r}{R})\left(\frac{c^2}{\gamma-1}+\frac{(v-v_1^{sh}\frac{r}{R})^2}{2}\right)=(v_n-v_1^{sh}\frac{r}{R})
(\frac{c^2}{\gamma(\gamma-1)}+\frac{(v-v_1^{sh}\frac{r}{R})^2}{2}).
\label{eq31a}
\ee
In the non-dimensional variables (\ref{eq24a}) we have from (\ref{eq31a})

\be
\left(\frac{4r}{5t}V-\frac{2r}{3t}\right)\left[\frac{Z}{\gamma-1}\frac{16r^2}{25t^2}
+\frac{1}{2}\left(\frac{4r}{5t}V-\frac{2r}{3t}\right)^2\right]
\label{eq32a}
\ee
$$=\left(\frac{4r}{5t}-\frac{2r}{3t}\right)
\left[\frac{Z}{\gamma(\gamma-1)}\frac{16r^2}{25t^2}
+\frac{1}{2}\left(\frac{4r}{5t}V-\frac{2r}{3t}\right)^2\right].
$$
This relation reduces to

\be
(V-\frac{5}{6})\left[\frac{Z}{\gamma-1}+(V-\frac{5}{6})^2\right]=
(1-\frac{5}{6})\left[\frac{Z}{\gamma(\gamma-1)}+(V-\frac{5}{6})^2\right],
\label{eq33a}
\ee
from what follows the first integral in the form

\be
Z=\frac{(\gamma-1)(1-V)(V-\frac{5}{6})^2}{2(V-\frac{5}{6}-\frac{1}{6\gamma})}.
\label{eq34a}
\ee
At the shock $r=R$, $\xi=1$, with $Z(1)$ and $V(1)$ from (\ref{eq26}), the first integral becomes an identity,
what confirms its correctness.
At existence of the first integral (\ref{eq34a}) we may use only two differential equations
(\ref{eq28a}) and (\ref{eq29a}) for finding a solution of the problem, like in the classical Sedov case.

\subsection{Exact solution for the expanding meduim}

Similar to the static background medium in the previous section (see \cite{sedov},\cite{sedov1}),
we look for an analytic solution of (\ref{eq28a}),(\ref{eq29a}), where
the variable $Z$ is determined by the algebraic relation (\ref{eq34a}).
Excluding $\frac{d\ln G}{d\ln\xi}$ from (\ref{eq28a}),(\ref{eq29a}), we obtain the equation

\be
(\gamma-1)\frac{dV}{d\ln\xi}-(1-V)\frac{d\ln Z}{d\ln\xi}=\frac{5}{2}-3\gamma V+V.
\label{eq35a}
\ee
From (\ref{eq34a}) we find
\be
\frac{d\ln Z}{d\ln\xi}=-\frac{1}{1-V}\,\frac{dV}{d\ln\xi}+\frac{2}{V-\frac{5}{6}}\,\frac{dV}{d\ln\xi}
-\frac{\gamma}{\gamma V-\frac{5\gamma}{6}-\frac{1}{6}}\,\frac{dV}{d\ln\xi}.
\label{eq36a}
\ee
After substituting in (\ref{eq35a}) we find the following equation for the variable $V$

\be
\left(\gamma+1-\frac{2}{6V-5}+\frac{\gamma-1}{6\gamma V-5\gamma
-1}\right)\frac{d V}{d\ln \xi}=\frac{5}{2}-3\gamma V+V.
\label{eq37a}
\ee
Using relations

$$\,\frac{2}{(6V-5)(\frac{5}{2}-3\gamma V+V)}=\frac{12}{(20-15\gamma)(6V-5)} $$
\be
\label{eq38a}
+\frac{2(3\gamma-1)}{(20-15\gamma)(\frac{5}{2}-3\gamma V+V)},
\ee
\be
\label{38b}
\frac{1}{(6\gamma V-5\gamma-1)(\frac{5}{2}-3\gamma V+V)}=\frac{1}{17\gamma-15\gamma^2+1}\,\frac{6\gamma}{6\gamma V-5\gamma-1}
\ee
$$
+\frac{1}{17\gamma-15\gamma^2+1}\,\frac{3\gamma-1}{\frac{5}{2}-3\gamma V+V},
$$
we obtain a solution of (\ref{eq37a}) in the form

\be
\label{39a}
 -\frac{2}{20-15\gamma}\ln(V-\frac{5}{6})
+\frac{\gamma-1}{17\gamma-15\gamma^2+1}\ln(V-\frac{5}{6}-\frac{1}{6\gamma})
\ee
$$
+\left[-\frac{\gamma+1}{3\gamma-1}-\frac{\gamma-1}{17\gamma-15\gamma^2+1}
+\frac{2}{20-15\gamma}\right]\ln\left(V-\frac{5}{6\gamma-2}\right)
=\ln\xi+{\rm const}.
$$
Using the boundary condition for $\xi=1$ from (\ref{eq26a}), we obtain the solution for $V(\xi)$ as

\be
\label{eq40a}
\left[(\gamma+1)(3V-\frac{5}{2})\right]^{\mu_1}
\left[\frac{\gamma+1}{\gamma-1}(6\gamma V-5\gamma-1)\right]^{\mu_2}
\left[6(\gamma+1)\frac{3\gamma V-V-\frac{5}{2}}{15\gamma^2+\gamma-22}\right]^{\mu_3}=\xi,
\ee
with

\be
\mu_1= \frac{2}{15\gamma-20},\,\,\, \mu_2=\frac{\gamma-1}{17\gamma-15\gamma^2+1},
\label{eq41a}
\ee
$$\mu_3=-\frac{\gamma+1}{3\gamma-1}-\frac{\gamma-1}{17\gamma-15\gamma^2+1}
+\frac{2}{20-15\gamma}.$$
For finding a solution for $G(\xi)$ we write the equations (\ref{eq28a}) and (\ref{eq37a}) in the form

\be
1-(1-V)\frac{d \ln G}{d V}=-(3V-\frac{5}{2}) \frac{d \ln\xi}{d V},
\label{eq42a}
\ee
\be
\frac{d \ln\xi}{d V}=\frac{\gamma+1-\frac{2}{6V-5}+\frac{\gamma-1}{6\gamma V-5\gamma-1)}}{\frac{5}{2}-V(3\gamma-1)}.
\label{eq43a}
\ee
The equation for $G(V)$ is obtained from these equations in the form

\be
1-(1-V)\frac{d \ln G}{d V}=-(3V-\frac{5}{2})\frac{\gamma+1-\frac{2}{6V-5}+
\frac{\gamma-1}{6\gamma V-5\gamma-1}}{\frac{5}{2}-V(3\gamma-1)}.
\label{eq44a}
\ee
It is reduces to

$$
(1-V)\frac{d \ln G}{d V}=-\frac{4}{3\gamma-1}+\frac{\frac{15\gamma}{2}-10}{3\gamma-1}\,
\frac{\gamma+1}{(3\gamma-1)V-\frac{5}{2}}
$$
\be
+\frac{1}{(3\gamma-1)V-\frac{5}{2}}
-\frac{\gamma-1}{6\gamma V-5\gamma-1}\,\frac{3V-\frac{5}{2}}{V(3\gamma-1)-\frac{5}{2}}.
\label{eq45a}
\ee
Here the relation was used:

\be
\label{45b}
\frac{3V-\frac{5}{3}}{\frac{5}{2}-3\gamma V +V}=\frac{1}{3\gamma-1}\left(-3+\frac{20-15\gamma}{2(\frac{5}{2}-3\gamma V +V)}\right).
\ee
Using algebraic relations (\ref{38b}),(\ref{45b}),

\be
\label{45c}
\frac{1}{(1-V)(6\gamma V-5\gamma-1)}=\frac{1}{\gamma-1}\left(\frac{1}{1-V}+\frac{6\gamma}{6\gamma V-5\gamma-1}\right),
\ee
\be
\label{48c}
\frac{1}{(1-V)(\frac{5}{2}-3\gamma V+V)}=\frac{1}{3\gamma-\frac{7}{2}}\left(-\frac{1}{1-V}+\frac{3\gamma-1}{\frac{5}{2}-3\gamma V+V}\right),
\ee
we obtain the solution of (\ref{eq45a}) in the form

\be
\ln G=\kappa_1\ln(1-V))+\kappa_2\ln(V-\frac{5\gamma+1}{6\gamma})
\label{eq49a}
+\kappa_3\ln(V-\frac{5}{6\gamma-2})+{\rm\bf const_1}.
\ee
Here
$$
\kappa_1=\frac{7}{3\gamma-1}-\frac{2}{6\gamma-7}+\frac{(15\gamma-20)(\gamma-1)}{(6\gamma-7)(15\gamma^2-17\gamma-1)}
$$
\be
\label{eq50a}
-\frac{3\gamma(15\gamma-20)}{(3\gamma-1)(15\gamma^2-17\gamma-1)}-\frac{15\gamma-20}{3\gamma-1}\,\frac{\gamma+1}{6\gamma-7},
\ee
$$
\kappa_2=-\frac{3}{3\gamma-1}+\frac{3\gamma(15\gamma-20)}{(3\gamma-1)(15\gamma^2-17\gamma-1)}.
$$
$$
\kappa_3=\frac{2}{6\gamma-7}-\frac{(15\gamma-20)(\gamma-1)}{(6\gamma-7)(15\gamma^2-17\gamma-1)}
+\frac{15\gamma-20}{3\gamma-1}\,\frac{\gamma+1}{6\gamma-7},
$$
The {\bf const$_1$} is found from the boundary conditions (\ref{eq26a}), and finally we obtain the solution for $G(V)$ in the form

\be
\label{eq51a}
G(V)=\frac{\gamma+1}{\gamma-1}\left[6\frac{(\gamma+1)(1-V)}{\gamma-1}\right]^{\kappa_1}
\left[\frac{\gamma+1}{\gamma-1}(6\gamma V-5\gamma-1)\right]^{\kappa_2}
\ee
$$
\times
\left(\frac{3(\gamma+1)}{15\gamma^2+\gamma-22}[(6\gamma-2)V-5)]\right)^{\kappa_3}.
$$
The function $Z(V)$ is determined by the integral (\ref{eq34a}).

\section{Discussion}

The constant $\beta$ in the definition of the non-dimensional radius $\xi$ in (\ref{eq25a}) is obtained from the
explosion energy integral $E$. The conserving value of the explosion energy behind the shock in
the uniformly expanding medium, with velocity and density distributions (\ref{eq20a}), is determined as

\be
\label{eq52a}
E=\int_0^{R(t)} \rho\left[\frac{\left(v-\frac{2r}{3t}\right)^2}{2}+\frac{c^2}{\gamma(\gamma-1)}\right]4\pi r^2 dr.
\ee
In non-dimensional variables (\ref{eq24a}) this relation reduces to the equation for the constant $\beta$

\be
\label{eq53a}
1=\beta^5\frac{64\pi}{25}\int_0^1 G\left[\frac{\left(V-\frac{5}{6}\right)^2}{2}+\frac{Z}{\gamma(\gamma-1)}\right]\xi^4 d\xi
\ee
The gravitational energy of the self-gravitation is not included directly in the explosion energy, but
implicitly it is present in the decreasing expansion velocity of the background solution (\ref{eq1b}).

It follows from the self-similar solution, that in the expanding medium the velocity of shock from (\ref{eq22a})
decreases as $\sim t^{-1/5}$, what is much slower than the shock velocity in the static uniform medium $\sim t^{-3/5}$, according
to Sedov solution (\ref{eq5}). Correspondingly the radius of the shock wave in the expanding self-gravitating medium
increases $\sim t^{4/5}$, more rapidly that the shock wave radius in the uniform non-gravitating medium $\sim t^{2/5}$.
It means, that the shock propagates in the direction of decreasing density with larger speed, than in the static medium,
 due to accelerating action
of the decreasing density, even in the presence of a self-gravitation.

\section*{Acknowledgments}

  This work  was partially supported by the Russian Federation President Grant
 for Support of Leading Scientific Schools, Grant No. NSh-261.2014.2,
  RFBR grant 14-02-00728, RAN Program 'Formation and evolution of stars and galaxies',
  and CAPES-ICRANet Programm.
    Author is grateful to A.D. Chernin for stimulating discussions, and
 Ivan Siutsou for help.


\begin{thebibliography}{100}


\bibitem{llhydro} Landau L. D., Lifshitz E. M. (1988) {\it Hydrodynamics}.
	Nauka, Moscow (in Russian)

\bibitem{llstat} Landau L. D., Lifshitz E. M. (1964) {\it Statistical Physics}.
	Nauka, Moscow (in Russian).

\bibitem{sedov} Sedov L.I. (1946) Doklady Acad. USSR {\bf 52}, No.1.

\bibitem{sedov1} Sedov L.I. (1977) {\it Metody podobiya i razmernostei v mekhanike}.
Nauka, Moscow (in Russian).

\bibitem{stanyuk}
Stanyukovich K.P. (1955) {\it Nonstationary motion of continuous media. Gostekhizdat}. Moscow (in Russian).

\bibitem{lzgrb} Tanvir N. (2013)  arXiv:1307.6156v1.

\bibitem{taylor} Taylor G.I. (1950), Proc. Roy. Soc. {\bf A201}, 175.

\bibitem{znuniv} Zeldovich Ya. B., Novikov I.D. (1983) {\it Relativistic astrophysics. Volume 2.
The structure and evolution of the universe}. Chicago, IL, University of Chicago Press.

\end{thebibliography}
\end{document}